# Room Temperature Magnetoelectric Effects on Single Slabs of Z-type Hexaferrites

Khabat Ebnabbasi [1], Yajie Chen[1], Anton Geiler[1], Vincent Harris[1], and Carmine Vittoria[1]

[1]Department of Electrical and Computer Engineering, Northeastern University, Boston, MA 02115, U.S.A.

## ABSTRACT

In this paper, magnetoelectric effects of Sr Z-type hexaferrite, $Sr_3Fe_{24}Co_2O_{41}$, at room temperature is measured. The change in remanence magnetization was measured by applying a DC voltage or electric field across a slab of hexaferrite. Changes of ~ 18% in remanence was observed in an electric field of 10,000V/cm implying a similar change in the microwave permeability at frequencies below 3GHz. In these types of measurements high resistivity is critical in order to reduce current flow in the hexaferrite. The resistivity of the hexaferrite was raised to $4.28 \times 10^8$ Ω-cm by annealing under oxygen pressure. The measurements indicate that indeed electric polarization and magnetization changes were induced by the application of magnetic and electric fields, respectively. The implications for microwave applications appear to be very promising at room temperature.

## INTRODUCTION

Since modern technologies will require miniaturization and efficient performances from the use of magnetic materials, inexpensive and simpler device structures must be developed in order to be compatible with the semiconductor technology. This may be achieved if all devices have the flexibility to be tuned by an electric field and/or voltage only- including ferrite devices. There have been a lot of efforts in the past decade to do away with magnetic fields and/or permanent magnets in the fabrication of microwave ferrite devices so that they may be tuned by an electric field or voltage. Multi-ferroic composite materials have been proposed to generate internal magnetic fields via voltage. Multi-ferroic composites usually consisted of a magnetostrictive and ferroelectric or piezoelectric slabs in physical contact whereby magnetic field sensors have been implied and fabricated so far. Also, small shifts in ferromagnetic resonance (FMR) have been observed using magnetoelectric composites in the presence of an electric field. To our knowledge tuning of ferrite microwave devices by an electric field or voltage is still not practical with present composite structures. We propose an alternative approach to this problem. A single layer of magneto-electric Z-type $Sr_3Co_2Fe_{24}O_{41}$ is proposed to induce magnetic parameter changes with application of voltage. The advantage of a single layer is that it is simpler to utilize to tune ferrite devices.

Hexagonal ferrites are of interest, since they exhibit high permeability at wireless frequencies [1,2]. In particular, $Co_2Z$-type ferrite, $Sr_3Co_2Fe_{24}O_{41}$, is a member of the planar hexaferrite family called *ferroxplana*, in which the easy magnetization direction lies in the basal plane (*c*-plane) of the hexagonal structure at room temperature. In this crystal, a large field is required to rotate the magnetic moments from the *c* plane, but a small field is enough for the moment in the c-plane. Hence, these materials are magnetically "soft" for H in the c-plane. As such the magnetic moments can follow an alternating field even in the gigahertz region, giving rise to high permeability even in the ultra high frequency (300 MHz–3 GHz) region. Therefore, this material is regarded as a promising candidate for inductor cores and electromagnetic noise absorbers to be used in this frequency region. Substitution of $Sr^{2+}$ for $Ba^{2+}$ was reported in order to reduce the sintering temperature from $1250(^0C)$ to $1210(^0C)$ and oxygen partial pressure in synthesizing $Co_2Z$-type ferrite [3]. This substitution also improved the frequency characteristic of the permeability. These results also indicate that $Sr^{2+}$ substitution would be favorable for lowering cost in manufacturing and putting this type of ferrite material into practical uses. Magnetic moments in $Ba_{1.5}Sr_{1.5}Co_2Fe_{24}O_{41}$ lie in the *c*-plane while that in $Sr_3Co_2Fe_{24}O_{41}$ are off the plane [4].

We report changes of remanence magnetization with the application of voltage in polycrystalline slabs of $Sr_3Co_2Fe_{24}O_{41}$ at room temperature as a manifestation of the magnetoelectric effect in this material. We deduce in this paper the implications of such changes to microwave applications.

**EXPERIMENTAL MATERIAL GROWTH PROCEDURE**

$Sr_3Fe_{24}Co_2O_{41}$ samples were prepared by solid-state reaction method [5]. The calculated amount of the oxide mixtures were: SrO (99.5%), $Co_3O_4$ (99.7%) and $Fe_2O_3$ (99.8%). 25gr of the starting reagents mixture were blended with a liquid dispersing agent (reagent alcohol). To grind uniformly, ball milling machine was used with a set of agate balls with 300rpm rotation speed for four hours. The slurry was dried at room temperature and five 5gr pellets were made. The pellets or discs were placed in the tube furnace in an oxygen atmosphere over the sample with 5deg/min temperature rate and set in 1210 ($^0C$) for 16 hours. To prevent formation of other impurity phases, including W- , M- or/and Y- phases, it was found most favorable to quench the sample immediately to room temperature. The X-ray diffraction pattern is shown in figure 1. In order to increase resistivity, samples were annealed at 600 ($^0C$) in an oxygen atmosphere for six hours [6]. The high resistivity of the sample was required for the magnetoelectric measurements to minimize current flow through the sample in the presence of high electric fields. The I-V curve of a typical sample is shown in figure 2, where the slab size was 1x1x0.5 $mm^3$. The current density was in the order of $10^{-3}$-$10^{-6}$ $A/cm^2$. Increasing the oxygen pressure during anneal reduced current flow as much as a factor of 100 and dependence of current with voltage is linear.

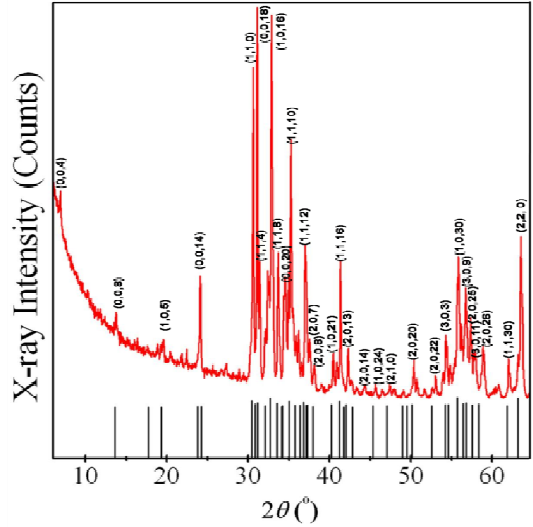

Figure 1. X-ray diffraction pattern of the polycrystalline $Sr_3Fe_{24}Co_2O_{41}$ at room temperature. The black line represents the reference peak positions for the Ba Z-type hexaferrite (Ref. ICDD # 19-0097. Space group: P63/mmc(194)).

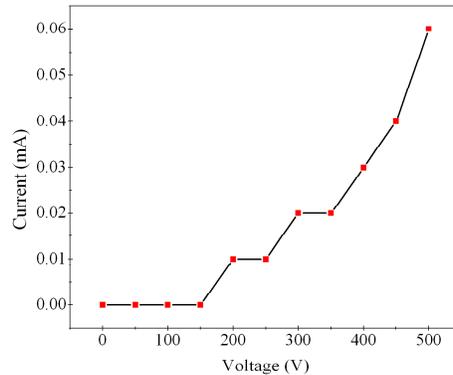

Figure 2. Polycrystalline Sr Z-type hexaferrite, $Sr_3Fe_{24}Co_2O_{41}$, I-V curve after increasing the resistivity.

**MAGNETO-ELECTRIC EXPERIMENTAL ANALYSIS**

In general terms, the ME effect implies the following: the application of a magnetic field, H, induces a change in electric polarization, P, and the application of an electric field, E, induces a change in magnetization, M. In figure 3(a), the magnetization, M, is plotted as a function of H, magnetic field, for a given application of electric field or voltage. We note that the remanence magnetization (for H=0) was indeed affected by voltage. The change in remanence magnetization was as much as 18% with the application of an electric field of 5KV/cm which is noticeable.

In figure 3(b) the percentage change in remanence versus the applied voltage is shown. Changes in remanence magnetization scale with polarity changes of the electric field or applied voltage. Thus, heating effects may be eliminated as a source of the remanence magnetization changes, since heating effects induce changes in remanence in one polarity sense only. The implication to microwave properties of this material is straightforward. The permeability expression for the Z or Y-type hexaferrite may be readily be found [7]. Typically ,the zero field FMR for these materials ranges near 3 GHz. However, below FMR frequency the permeability is approximately for this material to be:

$$\mu_r \approx 1+(4\pi M_R/ H_\varphi) \qquad (1)$$

where $M_R$ is the remanence magnetization and $H_\varphi$ is the c-plane magnetic anisotropy field. Typically, $H_\varphi$ is in the order of 40 Oe implying $\mu_r = 3.5$, since $4\pi M_R = 105G$. In figure 3(c) the permeability value versus frequency is given. Clearly, any changes in remanence magnetization is reflected in the microwave permeability at wireless communication frequencies. Certainly , application of DC voltages will not affect $H_\varphi$.

The mechanism for the ME effect is due to a local distortion of the Co ions giving rise to a spiral spin configuration which is a pre- required condition for this effect in hexaferrite materials. The electric field strains the material in a way to deform the spiral configuration and induce magnetic moment changes.

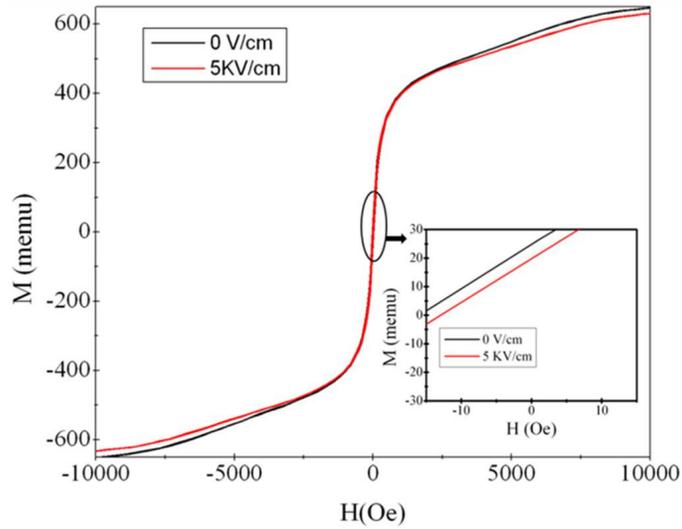

(a)

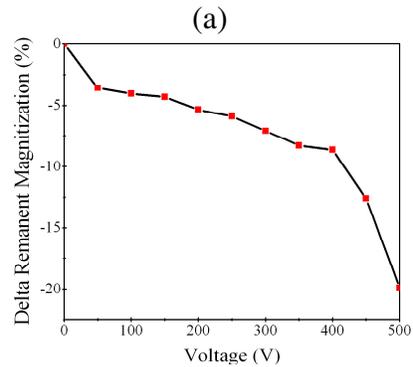

(b)

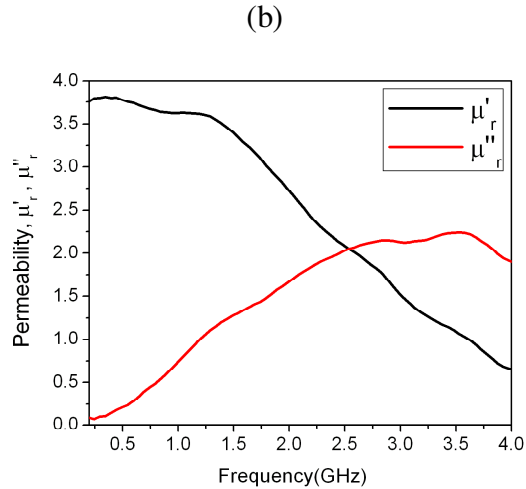

(c)

Figure 3. (a) Magnetization as a fucntion of H. The inset shows the effect of voltage at H=0. (b) Remanent magnitization shift vs different voltage values for typical sample with 0.5mm thickness.(c) Relative permeability versus microwave frequency.

## CONCLUSIONS

In this paper we reported changes of remanence magnetization with the application of voltage in polycrystalline slabs of $Sr_3Co_2Fe_{24}O_{41}$ at room temperature as a manifestation of the magnetoelectric effect in this hexaferrite material. Application of an electric field induces a strain which affects the spiral configuration and thus for the remanence magnetization. Work is in progress to present a direct way to measure permeability as a function of voltage at microwave frequency ranges as predicted in eq. 1.